\documentstyle[epsf]{article}

\newcommand{\ct}{\cite}
\newcommand{\lb}{\label}

\newcommand{\bc}{\begin{center}}
\newcommand{\ec}{\end{center}}
\newcommand{\bd}{\begin{displaymath}}
\newcommand{\ed}{\end{displaymath}}
\newcommand{\be}{\begin{equation}}
\newcommand{\ee}{\end{equation}}
\newcommand{\ba}{\begin{array}}
\newcommand{\ea}{\end{array}}
\newcommand{\bt}{\begin{tabular}}
\newcommand{\et}{\end{tabular}}
\newcommand{\un}{\underline}

\newcommand{\bp}{\begin{picture}}
\newcommand{\ep}{\end{picture}}
\newcommand{\bfi}{\begin{figure}}
\newcommand{\efi}{\end{figure}}

\def\fun#1#2{\lower3.6pt\vbox{\baselineskip0pt\lineskip.9pt
\ialign{$\mathsurround=0pt#1\hfil##\hfil$\crcr#2\crcr\sim\crcr}}}

\begin{document}

\title{
\huge\bf
{ Phase Transition in the Higgs Model of Scalar Fields with Electric
and Magnetic Charges }}
\author{\bf{ L.V.Laperashvili}\footnote{E-mails: laper@heron.itep.ru;
larisa@vxitep.itep.ru}\\
\it Institute of Theoretical and Experimental Physics,\\
\it B.Cheremushkinskaya 25, 117218 Moscow, Russia \\[0.2cm]
{\bf H.B.Nielsen}
\footnote{E-mails: hbech@nbivms.nbi.dk}\\
\it Niels Bohr Institute,\\
\it DK-2100, Copenhagen {\O}, Denmark}
\maketitle
\vspace*{0.1cm}
{\large
          renormalization group

\vspace{0.2cm}

\un{Corresponding author}:\\
Prof.H.B.Nielsen,\\
Niels Bohr Institute,\\
DK-2100, Copenhagen, Denmark\\
Tel: +45 35325259 \\
E-mail: hbech@alf.nbi.dk

\thispagestyle{empty}

\newpage

\thispagestyle{empty}

\vspace{10cm}
\begin{abstract}
Using a one-loop renormalization group improvement for the effective potential
in the Higgs model of electrodynamics with electrically and magnetically
charged scalar fields, we argue for the existence of a triple (critical)
point in the phase diagram ($\lambda_{run}, g_{run}^4$), where $\lambda_{run}$
is the renormalised running selfinteraction constant of the Higgs scalar
monopoles and $g_{run}$ is their running magnetic charge. This triple point
is a boundary point of three first-order phase transitions in the dual sector
of the Higgs scalar electrodynamics: The "Coulomb" and two
confinement phases meet together at this critical point.
Considering the arguments for the one-loop approximation validity in the
region of parameters around the triple point A we have obtained the
following triple point values of the running couplings:
$(\lambda_{(A)}, g^2_{(A)})\approx(-13.4; 18.6)$, which are independent of
the electric charge influence and two--loop corrections to $g^2_{run}$
with high accuracy of deviations. At the triple point the mass of monopoles
is equal to zero. The corresponding critical value of the electric
fine structure constant turns out to be $\alpha_{crit} =
\pi/g^2_{(A)}\approx{0.17}$ by the Dirac relation. This value is close to
the $\alpha_{crit}^{lat}\approx{0.20\pm 0.015}$, which in a $U(1)$ lattice
gauge theory corresponds to the phase transition between the "Coulomb"
and confinement phases. In our theory for $\alpha \ge \alpha_{crit}$
there are two phases for the confinement of the electrically charged particles.
The results of the present paper are very encouraging for the
Anti--grand unification theory which was developed previously as a realistic
alternative to SUSY GUTs. The paper is also devoted to the discussion of
this problem.
\end{abstract}
\newpage
\pagenumbering{arabic}

\vspace{1cm}

\normalsize

\section{Introduction}

\vspace{0.51cm}

Monte Carlo simulations of the lattice  U(1)--, SU(2)-- and  SU(3)-- gauge
theories \ct{1}-\ct{7} indicate the existence of a triple point on the
corresponding phase diagrams.
A triple point is a boundary point of three first order phase transitions.
Such a triple point is shown in Fig.1, which demonstrates the
results of the Monte Carlo simulations of the U(1) gauge theory described
by the following lattice action \ct{1}-\ct{3}:
\be
S = \sum_{\Box }[\beta\cos{\Theta (\Box )} + \gamma\cos{2\Theta (\Box )}].
                              \lb{1}
\ee
Here $\Theta (\Box )$ is a plaquette variable.

In the previous works \ct{8}-\ct{13} the investigation
of the phase transition phenomena and in particular, the
calculation of the U(1) critical coupling constant were connected with
the existence of artifact monopoles in the lattice gauge theory and also
in the Wilson loop action model, which we proposed in Ref.\ct{13}.

Now, instead of using the lattice or Wilson loop cut-off, we are going to
introduce physically existing monopoles into the theory as fundamental fields.
This idea was suggested previously in Refs.\ct{14}-\ct{16}.

Developing a version of the local field theory of the Higgs scalar
monopoles and electrically charged particles, we consider in Section 2
an Abelian gauge theory in the Zwanziger formalism \ct{17}--\ct{20}
and look for a/or rather several phase transitions connected with
the monopoles forming a condensate in the vacuum.
This Zwanziger formalism revealing a dual symmetry contains two
vector potentials $A_{\mu}(x)$ and $B_{\mu}(x)$
describing one physical photon. Using such an Abelian gauge theory
with both magnetic and electric charges,
we confirm in Section 3 rather simple expression for the effective
potential in the one-loop approximation which was obtained for the
Higgs scalar electrodynamics in Ref.\ct{21} (see also the review \ct{22}
and Ref.\ct{23}).
Using the renormalization group improvement of this effective potential
in Section 4, we investigate in Section 6 the phase structure of the
Higgs model with scalar monopoles.

In Section 5 we revise the results of Ref.\ct{20} where
the renormalization group equations for the electric and magnetic
fine structure constants were obtained in the case of
the existence of both charges. The Dirac relation
for the renormalized effective coupling constants is considered.
The influence of the electric charge on the monopole running fine
structure constant is investigated.

In Section 6 we show that the first--order phase transition
arises in the Higgs monopole model already on the level of
the "improved" one--loop approximation for the effective potential.
We argue for the existence of a triple point in the phase diagram
($\lambda_{run}; g^4_{run}$), where $\lambda_{run}$ is the renormalized
running selfinteraction constant in the Higgs monopole model and $g_{run}$
is the running magnetic charge of monopole.
We have obtained the triple point values $(\lambda_{(A)}; g^2_{(A)})
\approx(-13.4; 18.6)$ and the critical values of the magnetic fine structure
constant
$\tilde \alpha_{crit} = g^2_{(A)}/4\pi\approx{1.48}$
and electric fine structure constant
$\alpha_{crit} = \pi/g^2_{(A)}\approx{0.17}$ (by the Dirac relation).
The last value is very close to the lattice result \ct{3}:
$\alpha_{crit}^{lat}\approx{0.20\pm 0.015}$, which corresponds to the phase
transition between the "Coulomb" and confinement phases in the $U(1)$
lattice gauge theory.
Thus, if the lattice phase transition coupling roughly coincides
with our Coleman--Weinberg model it cannot depend much on lattice details.
Then we can expect that an "approximate universality"
(regularization independence) of the critical coupling constants,
previously suggested in Refs.\ct{10} and \ct{13}, takes place
even for the first order phase transitions. But this problem needs
further more exact investigations.

Section 7 is devoted to the description of the confinement of the
electrically charged particles by ANO "strings" --- electric vortices,
which exist in the phase with nonzero monopole Higgs field VEVs.

The results of the present paper are very encouraging for the Multiple
Point Model \ct{8}-\ct{16}, \ct{24}-\ct{28}. This problem is discussed
in Section 8.

\section{The Zwanziger Formalism for the Abelian Gauge Theory with Electric
and Magnetic Charges. Dual Symmetry}

A version of the local field theory of electrically and magnetically charged
particles is represented by Zwanziger formalism \ct{17},\ct{18}
(see also \ct{19} and \ct{20}), which considers
two potentials $A_{\mu}(x)$ and $B_{\mu}(x)$ describing one physical
photon with two physical degrees of freedom.
Now and below we call this theory QEMD ("Quantum
ElectroMagnetoDynamics").

In QEMD the total field system of the gauge, electrically ($\Psi$)
and magnetically ($\Phi$) charged fields (with charges $e$ and $g$,
respectively) is described by the partition
function which has the following form in Euclidean space:
\be
 Z = \int [DA][DB][D\Phi ][D\Phi^{+}][D\Psi ][D\Psi^{+}]e^{-S},
                                                         \lb{9}
\ee
where
\be
       S = \int d^4x L(x) =
           S_{Zw}(A,B) + S_{gf} + S_{(matter)}.
                                                       \lb{10}
\ee
The Zwanziger action $S_{Zw}(A,B)$ is given by:
$$
      S_{Zw}(A,B) = \int d^4x [\frac 12 {(n\cdot[\partial \wedge A])}^2 +
                  \frac 12 {(n\cdot[\partial \wedge B])}^2 +\\
$$
\be
     +\frac i2(n\cdot[\partial \wedge A])(n\cdot{[\partial \wedge B]}^*)
       - \frac i2(n\cdot[\partial \wedge B])(n\cdot{[\partial \wedge A]}^*)],
                                                         \lb{11}
\ee
where we have used the following designations:
$$
       {[A \wedge B]}_{\mu\nu} = A_{\mu}B_{\nu} - A_{\nu}B_{\mu},
\quad {(n\cdot[A \wedge B])}_{\mu} = n_{\nu}{(A \wedge B)}_{\nu\mu},\\
$$
\be
\quad {G}^*_{\mu\nu} = \frac 12\epsilon_{\mu\nu\lambda\rho}G_{\lambda\rho}.
                                           \lb{12}
\ee
In Eqs.(\ref{11}) and(\ref{12}) the unit vector $n_{\mu}$ represents
the fixed direction of the Dirac string in the 4--space.

The action $S_{(matter)}$:
\be
           S_{(matter)} = \int d^4x L_{(matter)}(x)         \lb{13}
\ee
describes the electrically and magnetically charged matter
fields.

$S_{gf}$ is the gauge--fixing action.

Let us consider now the Lagrangian $L_{(matter)}$ describing the Higgs
scalar fields $\Psi(x)$ and $\Phi(x)$ interacting with gauge fields
$A_{\mu}(x)$ and $B_{\mu}(x)$, respectively:
\be
   L_{(matter)}(x) = \frac 12 {|D_{\mu}\Psi|}^2 +
     \frac 12{|{\tilde D}_{\mu} \Phi |}^2 - U(\Psi, \Phi),  \lb{14}
\ee
where
\be
       D_{\mu} = \partial_{\mu} - ieA_{\mu},              \lb{15}
\ee
and
\be
       {\tilde D}_{\mu} = \partial_{\mu} - igB_{\mu}       \lb{16}
\ee
are covariant derivatives;
\be
U(\Psi, \Phi) = \frac 12 \mu_e^2{|\Psi|}^2 + \frac{\lambda_e}4{|\Psi|}^4
        + \frac 12 \mu_m^2{|\Phi|}^2 + \frac{\lambda_m}4{|\Phi|}^4
        + \lambda_1{|\Psi|}^2{|\Phi|}^2
                                                    \lb{17}
\ee
is the Higgs potential for the electrically and magnetically charged
fields $\Psi$ and $\Phi$.

The complex scalar fields:
\be
      \Psi = \psi + i\zeta \qquad \mbox{and}\qquad
      \Phi = \phi + i\chi                             \lb{18}
\ee
contain the Higgs $(\psi, \phi )$ and Goldstone $(\zeta, \chi )$ boson fields.

The gauge--fixed action chosen in Ref.\ct{19}:
\be
     S_{gf} = \int d^4x [\frac{M^2_A}2{(n\cdot A)}^2 +
                               \frac{M^2_B}2{(n\cdot B)}^2]
                                                     \lb{19}
\ee
has no ghosts.

The interactions in the Lagrangian (\ref{14}) are given by terms
$j_{\mu}^eA_{\mu}$ and $j_{\mu}^mB_{\mu}$
(here $j_{\mu}^e$ and $j_{\mu}^m$ are the electric and magnetic currents)
as well as by "sea--gull" terms $e^2A_{\mu}^2{|\Psi|}^2$ and
$g^2B_{\mu}^2{|\Phi|}^2$.
The local interaction of the electric and magnetic charges is described by
the term $\lambda_1{|\Psi|}^2{|\Phi|}^2$ of Eq.(\ref{17}) and also
is carried out via the "metamorphosic" propagator $<0|A_{\mu}B_{\nu}|0>$
(see Refs.\ct{19},\ct{20}).

Letting
\be
      F = \partial\wedge A = - (\partial\wedge B)^{*},        \lb{20}
\ee
\be
      F^{*} = \partial\wedge B = (\partial\wedge A)^{*},        \lb{21}
\ee
and the Hodge star operation (\ref{12}) on the field tensor:
\be
     F_{\mu\nu}^{*} = \frac 12 \epsilon_{\mu\nu\rho\sigma} F_{\rho\sigma},
                                                  \lb{22}
\ee
it is easy to see that the free Zwanziger Lagrangian (\ref{11})
is invariant under the following duality transformations:
\be
   F \leftrightarrow F^{*},\quad\quad (\partial\wedge A)
\leftrightarrow (\partial \wedge B),
   \quad\quad (\partial\wedge A)^{*} \leftrightarrow
         - (\partial \wedge B)^{*}.
                                                \lb{23}
\ee
The Lorentz invariance is lost in the Zwanziger Lagrangian (\ref{11})
because of depending on a fixed vector $n_{\mu}$, but this invariance
is regained for the quantized values of coupling constants $e$ and $g$
obeying the Dirac relation:
\be
                 e_ig_j = 2\pi n_{ij},~~~~~n_{ij} \in Z.
                                                      \lb{23a}
\ee
We also have a dual symmetry as the invariance of the total
Lagrangian $L(x)$ under the exchange of the electric and magnetic fields
(the Hodge star duality)
provided that at the same time the electric and magnetic charges and currents
transform according to the following discrete dual symmetry:
$$
   e \rightarrow g, \qquad
   g \rightarrow - e,
$$
\be
   \qquad j_{\mu}^e \rightarrow j_{\mu}^m,
   \qquad j_{\mu}^m \rightarrow - j_{\mu}^e.
                        \lb{24}
\ee
The masses and selfinteraction constants $\lambda_{e,m}$
of the electrically and magnetically charged particles
are also interchanged (if they are different):
\be
   \mu_e \leftrightarrow \mu_m,
   \qquad \lambda_e \leftrightarrow \lambda_m.     \lb{24a}
\ee
Considering the electric and magnetic fine structure constants:
\bc
\large{$\alpha = \frac {e^2}{4\pi} $ \quad \mbox{and} \quad
$\tilde \alpha = \frac {g^2}{4\pi} $ },
\ec
we have the invariance of the QEMD under the interchange
\be
          \alpha \leftrightarrow \tilde \alpha.       \lb{24b}
\ee
The case $n_{ij}=1$ corresponds to the Dirac relation for elementary
charges:
\be
              eg = 2\pi,                \lb{24c}
\ee
which can be rewritten in the following form:
\be
\alpha \tilde \alpha = \frac 14,    \lb{24d}
\ee
Then we elicit a fact of the invariance of the QEMD under the interchange:
\be
              \alpha \leftrightarrow \frac 1{4\alpha}.    \lb{24e}
\ee
Such a dual symmetry follows from Eq.(\ref{24}).
Now we are ready to calculate the effective potential.

\section{The Coleman--Weinberg Effective Potential for the Higgs Model
with Electrically and Magnetically Charged Scalar Fields}

The effective potential in the Higgs model of electrodynamics for
a charged scalar field was calculated in the one-loop approximation for the
first time by the authors of Ref.\ct{21}. The general method of the
calculation of the effective potential is given in the review \ct{22}.
Using this method we can construct the effective potential
(also in the one--loop approximation) for the theory described by the
partition function (\ref{9}) with the action $S$, containing the
Zwanziger action (\ref{11}),
gauge fixing action (\ref{19}) and the action (\ref{14})
for the electrically and magnetically charged matter fields.

Let us consider now the shifts:
\be
\Psi = \Psi_B + {\hat \Psi}(x), \quad\quad
 \Phi (x) = \Phi_B + {\hat \Phi}(x)
                                         \lb{25}
\ee
with $\Psi_B$ and $\Phi_B$ as background fields and calculate the
following expression for the partition function in the one-loop
approximation:
$$
  Z = \int [DA][DB][D\hat \Phi][D{\hat \Phi}^{+}][D\hat \Psi]
      [D{\hat \Psi}^{+}]\times \\
$$
$$
   \exp\{ - S(A,B,\Phi_B,\Psi_B)
   - \int d^4x [\frac{\delta S(\Phi)}{\delta \Phi(x)}|_{\Phi=
   \Phi_B}{\hat \Phi}(x) +
   \frac{\delta S(\Psi)}{\delta \Psi(x)}|_{\Psi=\Psi_B}{\hat \Psi}(x)
                + h.c. ]\}\\
$$
\be
    =\exp\{ - F(\Psi_B, \Phi_B, e^2, g^2, \mu_e^2, \mu_m^2,
                                \lambda_e, \lambda_m)\}.
                          \lb{26}
\ee

Using the representations (\ref{18}), we obtain the effective potential:
\be
  V_{eff} = F(\psi_B, \phi_B, e^2, g^2, \mu_e^2, \mu_m^2, \lambda_e,
                      \lambda_m)
                                        \lb{27}
\ee
given by the function $F$ of Eq.(\ref{26}) for the constant background
fields:
\be
\Psi_B = \psi_B = \mbox{const},\qquad\qquad
  \Phi_B = \phi_B = \mbox{const}.
                                                  \lb{28}
\ee
The effective potential (\ref{27}) has several minima. Their position
depends on $e^2, g^2, \mu^2_{e,m}$ and $\lambda_{e,m}$.
If the first local minimum occurs at $\psi_B=0$ and $\phi_B=0$,
it corresponds to the so-called "symmetrical phase", which is
the Coulomb-like phase in our description.

We are interested in the phase transition from the
Coulomb-like phase "$\psi_B = \phi_B = 0$" to the confinement phase
"$\psi_B = 0,\; \phi_B = \phi_0 \neq 0$". In this case the one--loop
effective potential for monopoles coincides with the expression
of the effective potential calculated by authors of Ref.\ct{21} for scalar
electrodynamics and extended to the massive theory in Ref.\ct{23}
(see review \ct{22}):
$$
 V_{eff}(\phi_B^2) = \frac{\mu_m^2}{2} {\phi_B}^2 +
                 \frac{\lambda_m}{4} {\phi_B}^4
$$
$$
  + \frac{1}{64\pi^2}[ 3g^4 {\phi_B}^4\log\frac{\phi_B^2}{M^2}
+ {(\mu_m^2 + 3\lambda_m {\phi_B}^2)}^2\log\frac{\mu_m^2 + 3\lambda_m
\phi_B^2}{M^2}
$$
\be
 + {(\mu_m^2 +\lambda_m \phi_B^2)}^2\log\frac{\mu_m^2
 + \lambda_m \phi_B^2}{M^2}] + C,
                           \lb{29}
\ee
where $M$ is the cut--off scale and C is a constant not depending on
$\phi_B^2$.

Considering the existence of the first vacuum at
$\phi_B=0$, we have $V_{eff}(0)=0$ and it is easy to determine
the constant C:
\be
        C = - \frac{\mu_m^4}{16{\pi^2}}\log \frac {\mu_m}M.  \lb{30}
\ee
Using from now the designations:$\quad\mu = \mu_m,\quad\lambda = \lambda_m$,
we have the effective potential in the Higgs monopole model described
by the following expression equivalent to Eq.(\ref{29}):
\be
V_{eff}(\phi^2)
= \frac{\mu^2_{run}}{2}\phi_B^2 + \frac{\lambda_{run}}{4}\phi_B^4
     + \frac{\mu^4}{64\pi^2}\log\frac{(\mu^2 + 3\lambda \phi_B^2)(\mu^2 +
       \lambda \phi_B^2)}{\mu^4}.
                                                \lb{31}
\ee
Here $\lambda_{run}$ is the running self--interaction constant
given by the expression standing before $\phi_B^4$ in Eq.(\ref{29}):
\be
  \lambda_{run}(\phi_B^2)
   = \lambda + \frac{1}{16\pi^2} [ 3g^4\log \frac{\phi_B^2}{M^2}
   + 9{\lambda}^2\log\frac{\mu^2 + 3\lambda \phi_B^2}{M^2} +
     {\lambda}^2\log\frac{\mu^2 + \lambda\phi_B^2}{M^2}].   \lb{32}
\ee
The running squared mass of Higgs scalar monopoles also follows from
Eq.(\ref{29}):
\be
   \mu^2_{run}(\phi_B^2)
   = \mu^2 + \frac{\lambda\mu^2}{16\pi^2}[ 3\log\frac{\mu^2 +
   3\lambda \phi_B^2}{M^2} + \log\frac{\mu^2 + \lambda\phi_B^2}{M^2}].
                                  \lb{33}
\ee
As it was shown in Ref.\ct{21}, the one--loop effective potential
(\ref{31}) can be improved by the consideration of the renormalization
group equation.

\section{The Renormalization Group Equation for the Improved Effective
Potential}

The renormalization group (RG) describes the dependence of a theory
and its couplings on an arbitrary scale parameter M. We are interested in
RG applied to the effective potential. In this case, knowing the dependence
on $M^2$ is equivalent to knowing the dependence on $\phi^2\equiv \phi_B^2$.
This dependence is given by the renormalization group equations (RGE).
Considering the RG improvement of the effective potential, we will
follow the approach of Coleman and Weinberg \ct{21} and its extension
to the massive theory \ct{23}, which are successively described in the
review \ct{22}. According to Refs.\ct{21}-\ct{23}, RGE for the
improved one--loop effective potential is given by the following
expression:
\be
 (M^2\frac{\partial}{\partial M^2} +
    \beta_{\lambda}\frac{\partial}{\partial \lambda} +
    \beta_g\frac{\partial}{\partial g} +
    \beta_{(\mu^2)}{\mu^2}\frac{\partial}{\partial \mu^2} -
    \gamma \phi^2 \frac{\partial}{\partial \phi^2}) V_{eff}(\phi^2) = 0,
				  \lb{34}
\ee
where the function $\gamma $ is the anomalous dimension:
\be
      \gamma(\frac{\phi}M) = - \frac{\partial \phi}{\partial M}.   \lb{35}
\ee
RGE (\ref{34}) leads to a new improved effective potential
(see the method of its obtaining in the review \ct{22}):
\be
   V_{eff}(\phi^2)
         = \frac 12 \mu^2_{run}(t) G^2(t)\phi^2 +
	       \frac 14 \lambda_{run}(t) G^4(t) \phi^4,   \lb{36}
\ee
where
\be
    G(t)\equiv \exp[ - \frac 12 \int_0^t dt' \gamma\biggl(g_{run}(t'),
	 \lambda_{run}(t')\biggl)].                \lb{37}
\ee
Eq.(\ref{34}) reproduces also a set of ordinary differential
equations:
\be
\frac{d\lambda_{run}}{dt} = \beta_{\lambda}\biggl(g_{run}(t),\lambda_{run}(t)
\biggl),
				      \lb{38}
\ee
\be
\frac{d\mu^2_{run}}{dt} = \mu^2_{run}(t) \beta_{(\mu^2)}\biggl(g_{run}(t),
      \lambda_{run}(t)\biggl),
				  \lb{39}
\ee
\be
\frac{dg_{run}}{dt} = \beta_g\biggl(g_{run}(t),\lambda_{run}(t)\biggl),
				   \lb{40}
\ee
where $t = \log (\phi^2/{M^2})$.

We can determine both beta functions for $\lambda_{run}$ and
$\mu^2_{run}$ by considering a change in M in the conventional
non--improved one--loop potential given by Eqs.(\ref{31}),
(\ref{32}) and (\ref{33}).

Let us write now the one--loop potential (\ref{31}) as
\be
	 V_{eff} = V_0 + V_1,                \lb{41}
\ee
where
\be
   V_0 = \frac{\mu^2}2 \phi^2 + \frac{\lambda}4 \phi^4  \lb{42}
\ee
and
$$
  V_1 = \frac{1}{64\pi^2}[ 3g^4 {\phi}^4\log\frac{\phi^2}{M^2}
+ {(\mu^2 + 3\lambda {\phi}^2)}^2\log\frac{\mu^2 + 3\lambda
\phi^2}{M^2}
$$
\be
   + {(\mu^2 +\lambda \phi^2)}^2\log\frac{\mu^2
+ \lambda \phi^2}{M^2} - 2\mu^4\log\frac{\mu^2}{M^2}].
                           \lb{43}
\ee
We can plug this $V_{eff}$ into RGE (\ref{34}) and obtain the
following equation (see \ct{22}):
\be
   ( \beta_{\lambda}\frac{\partial}{\partial \lambda} +
    \beta_{(\mu^2)}{\mu^2}\frac{\partial}{\partial \mu^2} -
    \gamma \phi^2 \frac{\partial}{\partial \phi^2}) V_0 =
     - M^2\frac{\partial V_1}{\partial M^2}.
				  \lb{44}
\ee
Equating $\phi^2$ and $\phi^4$ coefficients, we obtain:
\be
    \beta_{\lambda} = 2\gamma \lambda_{run}
                       + \frac{5\lambda_{run}^2}{8\pi^2} +
		\frac{3g_{run}^4}{16\pi^2},              \lb{45}
\ee
\be
    \beta_{(\mu^2)} = \gamma + \frac{\lambda_{run}}{4\pi^2}.    \lb{46}
\ee
The result for $\gamma$ is given in Ref.\ct{21} for scalar field with
electric charge $e$, but it is easy to rewrite this $\gamma$--expression
for monopoles with charge $g=g_{run}$:
\be
	  \gamma = - \frac{3g_{run}^2}{16\pi^2}.          \lb{47}
\ee
Finally we have:
\be
\frac{d\lambda_{run}}{dt} =
 \frac 1{16\pi^2}( 3g^4_{run} +10 \lambda^2_{run} - 6\lambda_{run}g^2_{run}),
				      \lb{48}
\ee
\be
\frac{d\mu^2_{run}}{dt} = \frac{\mu^2_{run}}{16\pi^2}( 4\lambda_{run} -
			   3g^2_{run} ).
				                       \lb{49}
\ee
Now the aim is to calculate the $\beta_g$--function, and here, as it was
shown in the next Section, we have the difference between the scalar
electrodynamics \ct{21} and scalar QEMD.

\section{Renormalization Group Equation for the Magnetic Charge.
The Dirac Relation}

It is well-known that in the absence of monopoles the Gell--Mann--Low equation
has the following form:
\be
     \frac{\mbox{d}(\log \alpha (p))}{\mbox{d}t} = \beta (\alpha (p)),
                          \lb{51}
\ee
where $t = \log(p^2/M^2)$, $p$ is a 4-momentum and $\alpha = e^2/4\pi$.
Let us consider also an energy scale $\tilde \mu$: $p^2 = {\tilde \mu}^2$
and $t = 2\log(\tilde \mu/M)$.

As it was first shown in Refs.\ct{29},
at sufficiently small charge ($\alpha < 1$) the $\beta-$function
is given by series over $\alpha /4\pi$:
\be
\beta (\alpha ) = \beta_2 (\frac{\alpha}{4\pi}) +
                                  \beta_4 {(\frac {\alpha}{4\pi})}^2 + ...
                                            \lb{52}
\ee
The first two terms of this series were calculated in the QED long time ago
in Refs.\ct{29},\ct{30}.
The following result was obtained in the framework of perturbation
theory (in the one- and two-loop approximations):

a)
\be
\beta_2 = \frac 43 \quad \mbox{and}\quad \beta_4 = 4 \quad
      -\quad \mbox{for fermion (electron) loops},
                                  \lb{53}
\ee
and

b)
\be
   \beta_2 = \frac 13 \quad \mbox{and}\quad \beta_4 =1 \quad
      -\quad \mbox{for scalar particle loops}.
                                          \lb{54}
\ee
This result means that for scalar particles
the $\beta$-function can be represented by the following series
aroused from Eq.(\ref{52}):
\be
  \beta(\alpha ) = \frac{\alpha}{12\pi}( 1
                               + 3 \frac{\alpha}{4\pi} + ...)
                                                   \lb{55}
\ee
and we are able to exploit the one-loop approximation (given by
the first term of Eqs.(\ref{52}) and (\ref{55}))
up to $\alpha \sim 1$ (with accuracy $\approx 25\%$ for
$\alpha \simeq 1$).

The Dirac relation and the renormalization group equations (RGE)
for the electric and magnetic fine structure constants $\alpha $
and $\tilde \alpha $ were investigated in detail by the authors in their
recent paper \ct{26}, where the same Zwanziger formalism was developed
for the QEMD. The following result was obtained.

If we have the electrically and magnetically charged particles
existing simultaneously for $\tilde \mu > \tilde \mu_{(threshold)}$
and if in some region of $\tilde \mu$ their Gell--Mann--Low
$\beta$--functions are computable perturbatively as a power series in $e^2$
and $g^2$, then the Dirac relation \ct{31} is valid not only for the "bare"
elementary charges $e_0$ and $g_0$, but also for the
renormalized effective charges $e$ and $g$ (see the history of this
problem in Refs.\ct{32} and in the review \ct{33}):
\be
	      eg = e_0g_0 = 2\pi.
                                     \lb{56}
\ee
Then
\be
        \alpha (t)\tilde{\alpha}(t) = \frac 14 \qquad
{\mbox{for any}}\quad \tilde \mu~ \ge~ \tilde \mu_{\mbox{(threshold)}}.
                                        \lb{57}
\ee
Eq.(\ref{57}) confirms the equality:
\be
 \frac{d\log \alpha(p)}{dt} = - \frac{d\log \tilde \alpha(p)}{dt}
					   \lb{58}
\ee
in the corresponding region of $t$.

Now if there exists a region of $\alpha $ and $\tilde \alpha$, where we
can describe monopoles and electrically charged scalar particles by a
perturbation theory, then the following RG--equations (obtained in \ct{20})
take place:
\be
  \frac{ d\log \alpha(p)}{dt} = -
  \frac{ d\log {\tilde \alpha}(p)}{dt} =
   \beta^{(e)}(\alpha ) - \beta^{(m)}(\tilde \alpha ). \lb{59}
\ee
These RGE are in accordance not only with the Dirac relation (\ref{57}),
but also with the dual symmetry considered in Section 2.

If both $\alpha $ and $\tilde{\alpha}$ are sufficiently small,
then the functions $\beta^{(e,m)}$ in Eq.(\ref{59}) are described by the
contributions of the electrically and magnetically charged particle
loops \ct{20}. Their analytical expressions are given by
the usual series similar to (\ref{52}).
By restricting ourselves to the two--loop approximation,
we have the following equations (\ref{59}) for scalar particles:
\be
  \frac{ d\log \alpha(p)}{dt} = -
  \frac{ d\log {\tilde \alpha}(p)}{dt} =
   \frac{\alpha - \tilde \alpha }{12\pi}( 1 + 3\frac{\alpha
          + \tilde \alpha}{4\pi} + ....).
                      \lb{60}
\ee
It is not difficult to see that $\beta^{(e,m)}$
coincide with the usual well--known $\beta$--functions
calculated in QED only in the two--loop approximation.
The deviation from the QED expressions for $\beta$--functions
arises only on the level of the higher order approximations
when the monopole (or electric particle) loops begin to play a role in the
electric (or monopole) loops.

According to Eq.(\ref{60}), the two--loop contribution is
not more than $30\%$ if both $\alpha $ and
$\tilde {\alpha}$ obey the following requirement:
\be
  0.25 \stackrel{<}{\sim }{\large \alpha, \tilde{\alpha}}\stackrel{<}
                {\sim } 1.
                                         \lb{61}
\ee
The lattice simulations of compact QED give the behavior of
the effective fine structure constant $\alpha(\beta)$
($\beta=1/e_0^2$ in Eq.(\ref{1}), and $e_0$ is the bare electric charge)
in the vicinity of the phase transition point
(see Refs.\ct{2},\ct{3},\ct{5}).
The following critical value of the fine structure
constant $\alpha $ was obtained in Ref.\ct{3}:
\be
\alpha_{crit}^{lat}\approx{0.20 \pm 0.015}\quad\quad
\mbox{at}\quad\quad \beta_{crit}\approx{1.011}.
                              \lb{62}
\ee
By the Dirac relation (\ref{57}), it is easy to obtain the corresponding
critical value of the monopole fine structure constant:
\be
    {\tilde \alpha}_{crit}^{lat}\approx{1.25 \pm 0.10}.    \lb{63}
\ee
These lattice critical values of the fine structure constants
$\alpha $ and $\tilde \alpha$
correspond to the phase transition point $(\beta, \gamma)\approx (1, 0)$
belonging to the phase diagram presented in Fig.1.

Eqs.(\ref{62}) and (\ref{63}) demonstrate that $\alpha $ and
$\tilde \alpha\;$ considered in the compact (lattice) QED in the vicinity
of the phase transition point almost coincide with the borders of the
requirement (\ref{61}) given by the perturbation theory for
$\beta$--functions.

Assuming that in the vicinity of the phase transition point
the coupling constant $g_{run}$ may be described by the one--loop
approximation, we obtain from Eq.(\ref{60})
the following RGE for $g_{run}^2$:
\be
 \frac{d\log(g_{run}^2)}{dt} = \frac{g^2_{run} - e^2_{run}}{48\pi^2}.
			      \lb{64}
\ee
Using the Dirac relation (\ref{57}), we have:
\be
\frac{dg^2_{run}}{dt} = \frac{g^4_{run}}{48\pi^2} - \frac 1{12}.
				   \lb{65}
\ee
Note that the second term of Eq.(\ref{65}) describes the influence of
the electrically charged fields on the behavior of the monopole charge.
In the case $\tilde \alpha >> \alpha$ we can neglect the second term on
the right side of Eq.(\ref{65}). Then we have the usual running coupling
in the dual sector of the scalar electrodynamics, which according
to Eq.(\ref{60}), is given by the following RGE in the
two--loop approximation:
\be
\frac{dg^2_{run}}{dt} = \frac{g^4_{run}}{48\pi^2}
                        + \frac{g^6_{run}}{{(16\pi^2)}^2}.
                                   \lb{65a}
\ee
These expressions of RGE for $g^2_{run}$ will be exploited in the next Section
for the investigation of the phase transition couplings in our model.

\section{Calculation of the Triple Point Couplings in the
Higgs Monopole Model of U(1) Gauge Theory}

As it was mentioned in Section 3, we consider the phase transition from
the Coulomb--like phase "$\psi_B = \phi_B = 0$" to the phase with
"$\psi_B = 0,\; \phi_B = \phi_0 \neq 0$". This means that the
effective potential (\ref{36}) of the Higgs scalar monopoles has
the first and the second minima appearing at $\phi = 0$ and
$\phi = \phi_0$, respectively. They are shown in Fig.2 by the curve "1".
These minima of $V_{eff}(\phi^2)$ correspond to the different vacua
arising in the model.
The conditions for the existence of degenerate vacua are given by the
following equations:
\be
          V_{eff}(0) = V_{eff}(\phi_0^2) = 0,     \lb{73}
\ee
\be
    \frac{\partial V_{eff}}{\partial \phi}|_{\phi=0} =
    \frac{\partial V_{eff}}{\partial \phi}|_{\phi=\phi_0} = 0
                                    \lb{74}
\ee
with inequalities
\be
 \frac{\partial^2 V_{eff}}{\partial \phi^2}|_{\phi=0} > 0,\quad
 \frac{\partial^2 V_{eff}}{\partial \phi^2}|_{\phi=\phi_0} > 0.
                                \lb{75}
\ee
The $\phi=\phi_0$ equation and inequality here are equivalent to:
\be
               V'_{eff}(\phi_0^2)
    \equiv \frac{\partial V_{eff}}{\partial \phi^2}|_{\phi=\phi_0} = 0,
                                    \lb{76}
\ee
\be
                V''_{eff}(\phi_0^2)
    \equiv \frac{\partial^2 V_{eff}}{\partial {(\phi^2)}^2}|_{\phi=\phi_0} > 0
                               \lb{77}
\ee
using $\phi^2$ as variable.
The first equation (\ref{73}) applied to Eq.(\ref{36}) gives:
\be
       \mu_{run}^2 = - \frac 12 \lambda_{run}(t_0){\phi_0^2}G^2(t_0),
					      \lb{78}
\ee
where $t_0 = \log(\phi_0^2/M^2)$.

The calculation of the first derivative of $V_{eff}(\phi^2)$ leads to the
following expression:
$$
 V'_{eff}(\phi^2) = \frac{V_{eff}(\phi^2)}{\phi^2}(1 + 2\frac{d\log G}{dt}) +
               \frac 12 \frac{d\mu^2_{run}}{dt} G^2(t)
$$
\be
  + \frac 14 \biggl(\lambda_{run}(t) + \frac{d\lambda_{run}}{dt} +
      2\lambda_{run}\frac{d\log G}{dt}\biggr)G^4(t)\phi^2.
					 \lb{79}
\ee
From Eq.(\ref{37}) and (\ref{47}) we have:
\be
    \frac{d\log G}{dt} = - \frac 12 \gamma = \frac{3g^2_{run}}{32\pi^2}.
					      \lb{80}
\ee
Using Eqs.(\ref{48}), (\ref{49}), (\ref{78}) and (\ref{80}), it is easy
to find the joint solution of equations
$V_{eff}(\phi_0^2) = V'_{eff}(\phi_0^2) = 0 $:
\be
    {V'}_{eff}(\phi_0^2) =
   (\frac 3{16\pi^2}g^4_{run} + \lambda_{run} + \frac 38 \frac
   {{\lambda_{run}}^2}{\pi^2})G^4(t_0)\phi_0^2 = 0,
                                                   \lb{81}
\ee
or
\be
     g^4_{crit} = - 2\lambda_{run}(\frac{8\pi^2}3 + \lambda_{run}). \lb{82}
\ee
The curve (\ref{82}) is represented on the phase diagram
$(\lambda_{run}; g^4_{run})$ of Fig.3 by the curve "1" which describes
a border between the "Coulomb--like" phase with $V_{eff} \ge 0$
and the confinement ones having $V_{eff}^{min} < 0$.

The next step is the calculation of the second derivative of the effective
potential:
$$
{V''}_{eff}(\phi^2) = \frac {{V'}_{eff}(\phi^2)}{\phi^2} + \biggl( - \frac 12
 \mu^2_{run}
+ \frac 12 \frac{d^2\mu^2_{run}}{dt^2} + 2\frac{d\mu^2_{run}}{dt}
 \frac{d\log G}{dt}
$$
$$
 +  \mu^2_{run}\frac{d^2\log G}{dt^2} +
   2\mu^2_{run}{(\frac{d\log G}{dt})}^2\biggl)\frac {G^2}{\phi^2} + \biggl(
   \frac 12 \frac{d\lambda_{run}}{dt} + \frac 14 \frac {d^2\lambda_{run}}
  {dt^2} + 2\frac{d\lambda_{run}}{dt}\frac{d\log G}{dt}
$$
\be
 + 2\lambda_{run}\frac{d\log G}{dt} + \lambda_{run}\frac{d^2\log G}{dt^2} +
     4\lambda_{run}{(\frac{d\log G}{dt})}^2\biggl) G^4(t).  \lb{83}
\ee
Let us consider now the case when this second derivative
changes its sign giving a maximum of $V_{eff}$ instead of the minimum
at $\phi^2 = \phi_0^2$. Such a possibility is shown in Fig.2 by
the dashed curve "2".

Now two additional minima at $\phi^2 = \phi_1^2$ and $\phi^2 = \phi_2^2$
appear in our theory. They correspond to two different confinement phases
related with the confinement of the electrically  charged particles.
If these two minima are degenerate, then we have the following
requirements:
\be
       V_{eff}(\phi_1^2) = V_{eff}(\phi_2^2) < 0    \lb{84}
\ee
and
\be
        {V'}_{eff}(\phi_1^2) = {V'}_{eff}(\phi_2^2) = 0,   \lb{85}
\ee
which describe the border between the confinement phases "conf.1" and "conf.2"
presented in Fig.3. This border is given as a curve "3" at the phase
diagram $(\lambda_{run}; g^4_{run})$ drawn in Fig.3. The curve "3"
meets the curve "1" at the triple point A (see Fig.3).
According to the illustration shown in Fig.2, it is obvious
that this triple point A is given by the following requirements:
\be
    V_{eff}(\phi_0^2) = V'_{eff}(\phi_0^2) = V''_{eff}(\phi_0^2) = 0.
                                         \lb{86}
\ee
In contrast to the requirements:
\be
       V_{eff}(\phi_0^2) = V'_{eff}(\phi_0^2) = 0,    \lb{86a}
\ee
giving the curve "1", let us consider now the joint solution of the following
equations:
\be
         V_{eff}(\phi_0^2) = V''_{eff}(\phi_0^2) = 0 .    \lb{86b}
\ee
It is easy to obtain this solution
using Eqs.(\ref{83}), (\ref{78}), (\ref{48}), (\ref{49}) and (\ref{65}):
\be
{\cal F}(\lambda_{run}, g^2_{run}) - \pi^2(\lambda_{run} - 2g^2_{run}) = 0,
                                                      \lb{87}
\ee
where
$$
{\cal F}(\lambda_{run}, g^2_{run}) = 5g_{run}^6 +
  24\pi^2g_{run}^4 + 12\lambda_{run}g_{run}^4 - 9\lambda_{run}^2g_{run}^2
$$
\be
  + 36\lambda_{run}^3 + 80\pi^2\lambda_{run}^2 + 64\pi^4\lambda_{run}.
                                                      \lb{88}
\ee
The dashed curve "2" of Fig.3 represents the solution of
Eq.(\ref{87}) which is equivalent to Eqs.(\ref{86b}). The curve "2"
is going very close to the maximum of the curve "1".
It is natural to assume that the position of the triple point A
coincides with this maximum and the corresponding deviation can be explained
by our approximate calculations. Taking into account such an assumption,
let us consider the border between the phase "conf.1" having the first
minimum at nonzero $\phi_0$  with $V_{eff}^{min} = c_1 < 0$
and the phase "conf.2 " which reveals two minima with the second minimum
being the deeper one and having $V_{eff}^{min}=c_2 < 0$.
This border (described by the curve "3" of Fig.3) was calculated in the
vicinity of the triple point A by means of Eqs.(\ref{84}) and (\ref{85})
with $\phi_1$ and $\phi_2$ represented as $ \phi_{1,2} = \phi_0 \pm \epsilon$
with $\epsilon << \phi_0$. The result of such calculations gives the
following expression for the curve "3":
\be
  g^4_{run} = \frac 52 ( 5\lambda_{run} + 8 \pi^2) \lambda_{run} + 8\pi^4.
                                  \lb{89}
\ee

The curve "3" meets the curve "1" at the triple point A.

The piece of the curve "1" to the left of the point A describes the border
between the "Coulomb--like" phase and the phase "conf.1". In the vicinity
of the triple point A the second derivative $V_{eff}''(\phi_0^2)$ changes
its sign leading
to the existence of the maximum at $\phi^2=\phi_0^2$, in correspondence with
the dashed curve "2" of Fig.2.  By this reason, the curve "1"
does not already describe a phase transition border up to the
next point B when the curve "2" again intersects the curve "1"
at $\lambda_{(B)}\approx - 12.24$.
This intersection (again giving $V''_{eff}(\phi_0^2) > 0$) occurs
supprisingly quickly.

The right piece of the curve "1" along to the right of the point B
separates the "Coulomb" phase and the phase "conf.2". But
between the points A and B the phase transition border is going
slightly upper the curve "1". This deviation is very small and can't be
distinguished on Fig.3.

The joint solution of equations (\ref{86})
leads to the joint solution of Eqs.(\ref{82}) and (\ref{87}).

It is necessary to note that only $V''_{eff}(\phi^2)$ contains
the derivative $dg^2_{run}/dt$. According to Eqs.(\ref{64}) and (\ref{65}),
the influence of the electric charge is described by the second term
of Eq.(\ref{87}).

With aim to investigate a position of the triple point,
let us consider the joint solution of Eqs.(\ref{86}) in the
following three cases:

\vspace{0.1cm}

a) The joint solution of Eqs.(\ref{82}) and
\be
  {\cal F}(\lambda_{run}, g^2_{run}) = 0                 \lb{93}
\ee
neglects the influence of the electrically charged fields.

\vspace{0.1cm}

b) The joint solution of Eqs.(\ref{82}) and (\ref{87})
takes into account the electric charge influence.
\vspace{0.1cm}

c) In the two--loop approximation for $dg^2_{run}/dt\;$  and in the absence
of the electric charge influence (see RGE (\ref{65a})),
it is necessary to consider the joint solution of Eq.(\ref{82}) and
the following equation:
\be
  {\cal F}(\lambda_{run}, g^2_{run}) + \frac 3{32\pi^2}g^8 = 0.
                                                            \lb{94}
\ee
These solutions were obtained numerically and gave the following triple point
values of $\lambda_{run}$ and $g^2_{run}$:
\be
    a) \qquad\lambda_{(A)}\approx{ - 13.4073},\quad
              g^2_{(A)}\approx{18.6070};              \lb{95}
\ee
\be
    b) \qquad\lambda_{(A)}\approx{ - 13.3696},\quad
              g^2_{(A)}\approx{18.6079};             \lb{96}
\ee
\be
    c) \qquad\lambda_{(A)}\approx{ - 13.4923},\quad
              g^2_{(A)}\approx{18.6044}.              \lb{97}
\ee
Such results show that the triple point
position ($\lambda_{(A)}; g^2_{(A)})$ is independent of the electric charge
influence and two--loop corrections to $dg^2_{run}/dt$ with accuracy of
deviations $ < 1\%$.

The numerical solution also demonstrates that the triple point A
exists in the very neighborhood of the maximum of the curve (\ref{82})
and its position is approximately given by the following values:
\be
     \lambda_{(A)}\approx - \frac{4\pi^2}3\approx -13.4,
                                                            \lb{98}
\ee
\be
         g^2_{(A)} = g^2_{crit}|_{\mbox{for}\;
                      \lambda_{run}=\lambda_{(A)}}
                 \approx \frac{4\sqrt{2}}3{\pi^2}\approx 18.6.
                                                                 \lb{99}
\ee
Let us conclude now what description of the phase diagram shown in Fig.3
we have finally.

There exists three phases in the dual sector of the Higgs
scalar electrodynamics : "Coulomb--like phase" and the confinement
phases "conf.1" and "conf.2".
The border "1", which is described by the curve (\ref{82}), separates
the "Coulomb--like phase" ($V_{eff} \ge 0$) and the confinement phases
($V_{eff}^{min}(\phi_0^2) < 0,\; V'_{eff}(\phi_0^2)=0$).
The curve "1" corresponds to the joint solution of the equations
$V_{eff}(\phi_0^2)=V'_{eff}(\phi_0^2)=0$ in the case b).

The dashed curve "2" represents the solution of the equations
$V_{eff}(\phi_0^2)=V''_{eff}(\phi_0^2)=0$.

The phase border "3" of Fig.3 separates two confinement phases.
The following requirements take place for this border:
$$
          V_{eff}(\phi_{1,2}^2) < 0,\qquad
$$
$$
         V_{eff}(\phi_1^2) = V_{eff}(\phi_2^2),
$$
$$
         V'_{eff}(\phi_1^2) = V'_{eff}(\phi_2^2) = 0,
$$
\be
         V''_{eff}(0) > 0,\qquad V''_{eff}(\phi_2^2) > 0.
                                                    \lb{100}
\ee
The triple point A is a boundary point of all three phase transitions
shown in the phase diagram of Fig.3.
For $g^2 < g^2_{({A})}$ the field system described by our model exists
in the confinement phase, where all electric charges  are confined.

The triple point value of the magnetic fine structure constant
follows from Eq.(\ref{99}):
\be
          \tilde \alpha_{(A)} = \frac {g^2_{(A)}}{4\pi}
                 \approx 1.48.          \lb{101}
\ee
By the Dirac relation (\ref{57}), we have calculated the value of the
triple point electric fine structure constant:
\be
      \alpha_{(A)} = \frac{\pi}{g^2_{(A)}}
                \approx{0.17}.
                                          \lb{102}
\ee
The obtained result is very close to the Monte Carlo lattice result
(\ref{62}).
Taking into account that monopole mass $m$ is given by the
following expression:
\be
  V''_{eff}(\phi_0^2) =
\frac 1{2\phi_0^2}\frac {d^2V_{eff}}{d\phi^2}|_{\phi=\phi_0}
            = \frac {m^2}{2\phi_0^2},                       \lb{103}
\ee
we see that monopoles acquire zero mass in the vicinity of the triple point A:
\be
  V''_{eff}(\phi_{0A}^2) = \frac {m^2_{(A)}}{2\phi_{0A}^2} = 0,
                               \lb{104}
\ee
and we can compare the values (\ref{101}) and (\ref{102}) with the
corresponding results obtained in Refs.[5,6] for the compact QED described by
the Villain action:
\be
   \alpha \approx 0.1836,\quad \tilde \alpha \approx 1.36,\quad
m^2\approx 0 \quad-\quad \mbox{in the vicinity of the critical point}.
                                                             \lb{105}
\ee
The phase diagram drawn in Fig.3 corresponds to the following
region of parameters:
\be
0.17 \stackrel{<}{\sim }{\large \alpha, \tilde{\alpha}}
\stackrel{<}{\sim }1.5,               \lb{105a}
\ee
where this diagram can be described by the one--loop
(renormalization group improved) effective potential.
According to Eq.(\ref{60}), in the region (\ref{105a})
the contribution of two loops is given by the accuracy of deviations
not more than $30\%$, therefore the perturbation theory works in this
region.

It is necessary to note that the estimation of $\beta_{\lambda}(\lambda,g^2)$
in RGE (\ref{38})
indicates a slow convergence of the series over $\lambda$. The comparison
of the expressions for the effective potentials $V_{eff}^{(1)}$
and $V_{eff}^{(2)}$ corresponding to the one--loop and two--loop
approximations in $\lambda$, respectively, leads to the following
relation:
\be
 V_{eff}^{(2)} \approx V_{eff}^{(1)}(1 + \frac{a\lambda_{run}}{48\pi^2} + ...)
                                       \lb{106}
\ee
with $a\sim 1$. Such a result allows to consider the one--loop approximation
for $\lambda_{run}$ up to $|\lambda|\stackrel{<}{\sim }30$ with accuracy
of deviations $<10\%$, and the value $\lambda_{(A)}\approx - 13$
obtained in the present paper for the triple point A is expected as
well described by our method.
The analysis of such an estimation was performed with help of
Refs.\ct{34},\ct{35} applied to our case of the scalar Higgs fields.

\section {"ANO--strings", or the vortex description of the
confinement phases }

As it was shown in the previous Section, two regions between the curves
"1", "3" and "3", "1", given by the phase diagram of Fig.3, correspond to the
existence of two confinement phases, different in the sense
that the phase "conf.1" is produced by
the second minimum, but the phase "conf.2" corresponds to the third minimum
of the effective potential. It is obvious that in our case
both phases have nonzero monopole condensate in the minima of the effective
potential, when $V_{eff}^{min}(\phi_{1,2}\neq 0) < 0$. By this reason, the
Abrikosov--Nielsen--Olesen (ANO) electric vortices
(see Refs.\ct{36},\ct{37}) may exist in these both phases, which are
equivalent in the sense of the "string" formation.  If electric
charges are present in a model (they are given in our model by the
electrically charged Higgs field $\Psi$), then these charges are placed
at the ends of the vortices--"strings" and therefore are confined.

Utilizing the string formulation of our Abelian Higgs Model (AHM),
we can use the result of Ref.\ct{38} and consider the partition function:
\be
    Z_{AHM} = \int DB D\Phi D{\bar{\Phi}} e^{ - S_{AHM}(B,\Phi,\bar{\Phi}) }
                                                               \lb{107x}
\ee
with
\be
S_{AHM}(B,\Phi,\bar{\Phi}) = \int d^4x\{\frac 14 {(\partial \wedge B)}^2
         + \frac 12 {|(\partial - ig B)\Phi |}^2  +
         \frac {\lambda}4 {({|\Phi|}^2 - \Phi_0^2 )}^2\},
                                     \lb{108x}
\ee
which in the London limit $(\lambda \to \infty)$ can be rewritten in terms
of the world--sheet coordinates $X^{\mu}(\sigma)$ of the ANO closed strings:
\be
       \lim_{\lambda\to \infty} Z_{AHM} = \int_{\delta \Sigma = 0}
                         D\Sigma e^{-S(\Sigma)}.
                                      \lb{109x}
\ee
In Eq.(\ref{109x}) we have:
$$
         S(\Sigma) = \frac {\pi^2 m_V^2}{g^2}\int d^2\sigma d^2\sigma'
    \epsilon^{ab}\partial_a X^{\mu}(\sigma)\partial_b X^{\nu}(\sigma)
    K[X(\sigma) - X(\sigma')]
$$
\be
        \epsilon^{a'b'}\partial_{a'}
    X^{\mu}(\sigma')\partial_{b'} X^{\nu}(\sigma'),
                                      \lb{110x}
\ee
where $m_V^2 = g^2 \phi_0^2$. The kernel $K(x)$ obeys the following
equation:
\be
             (-{\partial}^2 + m_V^2) K (x) = \delta (x).
                                       \lb{111x}
\ee
As it is well--known \ct{36},\ct{37}, in the London's limit
($\lambda \to \infty$) the Abelian Higgs model, described by the Lagrangian
(\ref{14})--(\ref{17}) with $\Psi = 0$, gives the formation of
the condensate with amplitude $\phi_0$ which repels and suppresses
the electromagnetic field $F_{\mu\nu}$ almost everywhere, except the
region around the vortex lines. In this limit, we have the following
London equation:
\be
                {\mbox rot}\; {\vec j}^m = {\delta}^{-2}{\vec E},      \lb{112x}
\ee
where ${\vec j}^m$ is the microscopic current of monopoles,
${\vec E}$ is the electric field strength and $\delta$ is the penetration
depth. It is clear that ${\delta}^{-1}$ is the photon mass $m_V$,
generated by the Higgs mechanism. The closed equation for $\vec E$
follows from the Maxwell equations and Eq.(\ref{112x}) just in the
London's limit.

In our case $\delta$ is defined by the following relation:
\be
            {\delta}^{-2} \equiv m_V^2 = g^2 \phi_0^2.    \lb{113x}
\ee
On the other hand, the field $\phi$ has its own correlation length
$\xi$, connected to the mass of the field $\phi$ ("the Higgs mass"):
\be
        \xi = {m_S}^{-1}, \qquad m_S^2 = \lambda \phi_0^2.   \lb{114x}
\ee
The London's limit for our "dual superconductor of the second kind"
corresponds to the following relations:
\be
      \delta >> \xi,\qquad m_V << m_S,\qquad g << \lambda,    \lb{115x}
\ee
and "the string tension" --- the vortex energy per unit length (see
Ref.[37]) --- for the minimal electric vortex flux $2\pi$, is:
\be
       \sigma = \frac {2\pi}{g^2\delta^2}\ln \frac{\delta}{\xi}
              = 2\pi \phi_0^2 \ln \frac{m_S}{m_V}, \qquad {\mbox
        {where}}\qquad \delta/\xi = \frac {m_S}{m_V} >> 1.      \lb{116x}
\ee
We see that in the London's limit ANO--theory implies the mass generation
of the photons, $m_V = 1/\delta$, which is much less than the Higgs
mass $m_S = 1/\xi$.

Let us be interested now in the question whether our "strings" are
thin or not. The vortex may be considered as thin, if the distance
between the electric charges sitting at its ends, i.e. the string length $L$,
is much larger than the penetration length $\delta$:
\be
            L >> \delta >> \xi.         \lb{117x}
\ee
In the framework of classical calculations, it is not difficult to
obtain the mass $M$ and angular momentum $J$ of the rotating "string":
\be
             J = \frac{1}{2\pi \sigma} M^2, \qquad
             M = \frac {\pi}{2} \sigma L.               \lb{118x}
\ee
The following relation follows from Eqs.(\ref{118x}):
\be
        L = 2 \sqrt{\frac{2J}{\pi \sigma}},             \lb{119x}
\ee
or
\be
       L = \frac{2g\delta}{\pi}\sqrt{\frac{J}{\ln \frac{m_S}{m_V}}}.
                             \lb{120x}
\ee
For $J=1$ we have:
\be
       \frac{L}{\delta} = \frac{2g}{\pi \sqrt{\ln \frac {m_S}{m_V}}},
                                                            \lb{121x}
\ee
what means that for $ m_S >> m_V $ the length of this "string" is small
and does not obey the requirement (\ref{117x}). It is easy to see from
Eq.(\ref{120x}) that in the London's limit the "strings" are very thin
($L/\delta >> 1$)
only for the enormously large angular momenta $J >> 1$.

The phase diagram of Fig.3 shows the existence of the confinement phase
for $\alpha \ge \alpha_{(A)}\approx 0.17$. This means that the formation
of vortices begins at the triple point $\alpha = \alpha_{(A)}$:
for $\alpha > \alpha_{(A)}$ we have nonzero $\phi_0$ leading to the
existence of vortices.

It is necessary to emphasize that this value $\alpha_{(A)}$, as the
lattice $\alpha_{crit}$, is sufficiently small and corresponds to
the validity of the perturbation theory for $\beta$--functions of
RGE.

The lattice investigations show that in the confinement phase
$\alpha (\beta)$ increases when $\beta = 1/e_0^2 \to 0$
(here $e_0$ is the the bare electric charge) and very slowly
approaches to its maximal value (see Refs.[2,3]).
Such a phenomenon leads to the "freezing" of the electric fine
structure constant at the value $\alpha = \alpha_{max}$ due to the
Casimir effect.
The authors of Ref.\ct{39} predicted: $\alpha_{max}=\frac {\pi}{12}
\approx 0.26$ (see also Ref.\ct{13}).

Let us consider now the region of the confinement values of the magnetic
charge $g$ (obtained in this paper):
$$
               g_{min} \le g \le g_{max},
$$
$$
           g_{max} = g_{(A)}\approx{\sqrt {18.6}}\approx 4.3,
$$
\be
           g_{min} = \sqrt {\frac {\pi}{\alpha_{max}}}\approx 3.5.
                                             \lb{122x}
\ee
Then for $m_S = 10m_V$ say we have from Eq.(113) the following estimation
of the "string" length for $J=1$:
\be
            1.5 \stackrel{<}{\sim}\frac{L}{\delta}\stackrel{<}{\sim} 2.
                              \lb{123x}
\ee
We see that "low-lying" states of "strings" correspond to the short
and thick vortices.

It is worthwhile mentioning that the confinement of monopoles
is described by the non--dual (usual) sector of the Higgs electrodynamics
and the Higgs field $\Psi$, having the electric charge, is responsible
for this confinement.
The corresponding confinement phases for monopoles are absent
on the phase diagram of Fig.3. They can be described by the phase
diagram ($\lambda^e_{run}; e^2_{run}$). Of course, the dual symmetry predicts
that the triple point $e^2_{(A)}$ has to be given by the same
value (\ref{99}), i.e.
$\alpha_{(A)}\approx 1.48$ and $\tilde \alpha_{(A)}\approx 0.17$.
The overall phase diagram is three-dimensional and is given by
$(\lambda^m_{run}; \lambda^e_{run}; g^2_{run})$, because $e^2_{run}$ and
$g^2_{run}$ are related by the Dirac charge quantization condition.
It is natural to expect that
the region of the monopole confinement also stretches from
$\tilde \alpha = {\tilde \alpha}_{(A)}\approx 0.17$ up to the
value $\tilde \alpha = {\tilde \alpha}_{max}\approx 0.26$.

\section{Multiple Point Principle and the Higgs Monopole Model}

Most efforts to explain the Standard Model (SM) describing well all
experimental results known today are devoted to Grand Unification
Theories (GUTs). The supersymmetric extension of the SM consists of taking the
SM and adding the corresponding supersymmetric partners \ct{40}.  The Minimal
Supersymmetric Standard Model (MSSM) shows the possibility of the existence of
the grand unification point at $\mu_{GUT}\sim 10^{16}$ GeV \ct{41}.
But the absence of supersymmetric particle production at current accelerators
and additional constraints arisen from limits on the contributions of virtual
supersymmetric particle exchange to a variety of the SM processes indicate that
at present there are no unambiguous experimental results requiring the
existence of the supersymmetry \ct{42}, \ct{43}.

Anti--Grand Unification Theory (AGUT) was developed
in Refs.\ct{24}-\ct{28} as a realistic alternative to SUSY GUTs.
According to the AGUT, the supersymmetry does not come into the existence
up to the Planck energy scale:
\be
         \mu_{Pl} = 1.22\cdot 10^{19}\;\mbox{GeV}.    \lb{1}
\ee
The AGUT suggests that at the Planck scale $\mu_{Pl}$, considered as a
fundamental scale, there exists the more fundamental gauge group $G$,
containing three copies of the Standard Model Group ($SMG$) \ct{24}-\ct{28}:
\be
SMG = S(U(2)\times U(3)) = \frac{U(1)\times SU(2)\times SU(3)}
     {\{{(2\pi, -1^{2\times 2}, e^{i2\pi/3}1^{3\times 3})}^n|n\in
      Z\}},                                                       \lb{2}
\ee
\be
          G=(SMG)^3=SMG_1\otimes SMG_2\otimes SMG_3 .             \lb{3}
\ee
The fitting of fermion masses [28] suggests the generalized G:
\be
   G_f = {(SMG)}^3\otimes U(1)_f.     \lb{4}
\ee
The AGUT approach is used in conjunction with the Multiple Point Principle
(MPP) proposed several years ago in Refs.\ct{8}-\ct{10}.
According to this principle, Nature seeks a special point -- the Multiple
Critical  Point (MCP) where many phases meet.

In the AGUT the group $G$ undergoes (an order of magnitude under the Planck
scale) spontaneous breakdown to the diagonal subgroup:
\be
G\to G_{diag.subgr.}=\left\{g, g, g\parallel g\in SMG\right\}, \lb{5}
\ee
which is identified with the usual (low--energy) group SMG.

The MCP is a point on the phase diagram of the fundamental
regularized gauge theory G, where the vacua of all fields existing in
Nature are degenerate (MPP).

The idea of the MPP has its origin in the lattice investigations of
gauge theories \ct{1}-\ct{7}.

The precision of the LEP--data allows to extrapolate three running constants
$\alpha_{i} (\mu)$ of the SM (i=1,2,3 for U(1), SU(2), SU(3) groups)
to high energies with small errors and we are able to perform some
checks of GUTs and AGUT. The MPP predicts the following values of the
fine structure constants at the Planck scale in terms of the phase
transition couplings (see Refs.\ct{8}-\ct{10}):
\be
 \alpha_i(\mu_{Pl}) = \frac{\alpha_i^{crit}}{N_{gen}}
                  = \frac{\alpha_i^{crit}}{3}
           \qquad\mbox{for} \qquad i=2,3,                     \lb{7}
\ee
and
\be
 \alpha_1(\mu_{Pl}) = \frac{\alpha_1^{crit}}{{\frac 12}N_{gen}(N_{gen}+1)}
                    = \frac{\alpha_1^{crit}}{6}\qquad\mbox{for}\quad U(1),
                                                           \lb{8}
\ee
where $N_{gen}=3$ is the number of quark and lepton generations.

Eqs.(\ref{7}) and (\ref{8}) contain the phase transition values
$\alpha_i^{crit}$ of the fine structure constants $\alpha_i$.
This means that at the Planck scale the running constants $\alpha_1$ (or
$\alpha_Y\equiv\frac{3}{5}\alpha_1$), $\alpha_2$ and $\alpha_3$,
as chosen by Nature, are just the ones corresponding to the MCP.

Multiple Point Model (MPM) \ct{8}-\ct{16}, \ct{24}-\ct{28} assumes the
existence of the MCP at the Planck scale $\mu_{Pl}$. The MCP is a boundary
point of a number of the first order phase transitions in the system of all
fields presented by Nature beyond the SM.  We assume that the Higgs scalar
fields with dual charges (in particular, Higgs scalar monopoles of U(1)
gauge theory) are responsible for such phase transitions.

The extrapolation of the experimental values of the inverses
$\alpha^{-1}_{Y,2,3}(\mu)$ to the Planck scale $\mu_{Pl}$ by the
renormalization group formulas (in doing the extrapolation with one
Higgs doublet under the assumption of a "desert") leads to the
following result:
\be
\alpha^{-1}_Y(\mu_{Pl})=55.5;\quad\alpha^{-1}_2(\mu_{Pl})=49.5;
         \quad\alpha^{-1}_3(\mu_{Pl})=54.        \lb{107}
\ee
Eq.(\ref{8}) applied to the first value of Eq.(\ref{107}) gives the MPM
prediction for the U(1) fine structure constant at the phase transition
point (see details in \ct{10}):
\be
              \alpha_{crit}^{-1}\approx 8.    \lb{108}
\ee

The result (\ref{102}) obtained in this paper in the Higgs scalar monopole
model gives the following prediction:
\be
           \alpha_{crit}^{-1} = {\alpha_{(A)}}^{-1}
                    \approx 6,                               \lb{109}
\ee
which is comparable with the MPM result (\ref{108}).

Although the one--loop approximation for the (improved) effective potential
does not give an exact coincidence
with the MPM prediction of the critical $\alpha$,
we see that, in general, the Higgs monopole model is very
encouraging for the AGUT--MPM. We have a hope that the two--loop
approximation corrections to the Coleman--Weinberg effective potential will
lead to the better accuracy in calculation of the phase transition couplings.
But this is an aim of the next papers.

\section{Conclusions}

We have used the Coleman--Weinberg effective potential for the Higgs model
with the Higgs field $\phi$ conceived as a monopole scalar field
to enumerate a phase diagram suggesting that in addition to the phase with
$<\phi>=0$ (i.e. the Coulomb phase) we have two different
phases with $<\phi>\neq 0$ meaning confinement phases: "conf.1" and
"conf.2". These three phases
meet in the dual phase diagram at a triple point A and we calculated
the running $\lambda_{run}$ and $g^2_{run}$ couplings at this point:
\be
  \bigl(\lambda_{(A)}; {g^2_{(A)}}\bigr)\approx ( - \frac 43\pi^2;
\frac{4\sqrt{2}}{3}\pi^2)\approx{(-13.4; 18.6)}.  \lb{110}
\ee
By the Dirac relation, the obtained $g^2_{(A)}$ corresponds to
\be
     \alpha_{(A)} = \frac{\pi}{g^2_{(A)}}
      \approx{0.17}.                         \lb{111}
\ee
It is noticed that these triple point fine structure constant values
are very close to the phase transition values of the fine structure
constants given by a U(1) lattice gauge theory and Wilson loop action
model \ct{13}.

The review of all existing results gives:

1)
\be
    \alpha_{crit}^{lat}\approx{0.20 \pm 0.015},\quad
    {\tilde \alpha}_{crit}^{lat}\approx{1.25 \pm 0.10}     \lb{112}
\ee
-- in the Compact QED with the Wilson lattice action \ct{3};

2)
\be
    \alpha_{crit}^{lat}\approx{0.204} \quad
    {\tilde \alpha}_{crit}^{lat}\approx{1.25}    \lb{113}
\ee
-- in the model with the Wilson loop action \ct{13};

3)
\be
   \alpha_{crit} \approx 0.1836,\quad \tilde \alpha_{crit} \approx 1.36
                                                  \lb{114}
\ee
-- in the Compact QED with the Villain lattice action \ct{6};

4)
\be
     \alpha_{crit} = \alpha_{(A)}\approx{0.17},\quad
     {\tilde \alpha}_{crit} = {\tilde \alpha}_{(A)}\approx 1.48
                                     \lb{115}
\ee
-- in the Higgs scalar monopole model (the present paper).

\vspace{0.1cm}
It is necessary to emphasize that the functions $\alpha (\beta)$
for the effective electric fine structure constant are different
for the Wilson and Villain lattice actions in the U(1) lattice gauge theory,
but the critical values of $\alpha$ coincide for both theories
\ct{2},\ct{3}.

Hereby we see an additional arguments for our previously hoped (see
\ct{10} and \ct{13}) "approximate universality" of the first order
phase transition couplings: the fine structure constant (in the continuum) is
at the/a multiple point approximately the same one independent of various
parameters of the different (lattice, etc.) regularization.

All critical values (\ref{112})-(\ref{115})
correspond to the perturbative region of parameters:
\be
        0.17 \stackrel{<}{\sim }{\large \alpha, \tilde{\alpha}}
          \stackrel{<}{\sim }1.5,
                                       \lb{116}
\ee
when the two-loop contributions to RGE are $< 30\%$.

We could also comment:

All different versions of U(1) lattice gauge theories have
artifact monopoles. If they are approximated by a continuum field
model it should be the Higgs model interpreted as in the present article
and our triple point $\alpha_{(A)}$ would be the critical
(and maybe the triple point) coupling
of $\un{whatever}$ U(1) lattice gauge theory. This is our previously
suggested "approximate universality" which is very needed
for the AGUT and MPP predictions. To the point, the result
$\alpha_{(A)}\approx 0.17$ obtained in our Higgs monopole model
leads to $\alpha_{crit}^{-1}\approx 6$ which is comparable
with AGUT--MPP prediction $\alpha_{crit}^{-1}\approx 8$.
The details of this problem are discussed in Refs.\ct{8}-\ct{10}.

The results of the present paper stimulate the further investigations
of the phase transition phenomena in the Higgs model of scalar monopoles
with aim to obtain better accuracy for the phase transition coupling
values.

\vspace*{0.5cm}

ACKNOWLEDGMENTS: We would like to express a special thanks to
D.L.Bennett for useful discussions and D.A.Ryzhikh for
numerical calculations. We are also very thankful with
Colin Froggatt, Roman Nevzorov and
all participants of the Bled Workshop-2000 "What comes
beyond the Standard Model" (Bled, Slovenia, 17-28 July, 2000)
for stimulating interactions.

One of the authors (L.V.L.) thanks very much the Niels Bohr Institute
for its hospitality and financial support.

Also the financial support from grants INTAS-93-3316-ext and
INTAS-RFBR-96-0567 is gratefully acknowledged.

\newpage
\begin{figure}
\centerline{\epsfxsize=\textwidth \epsfbox{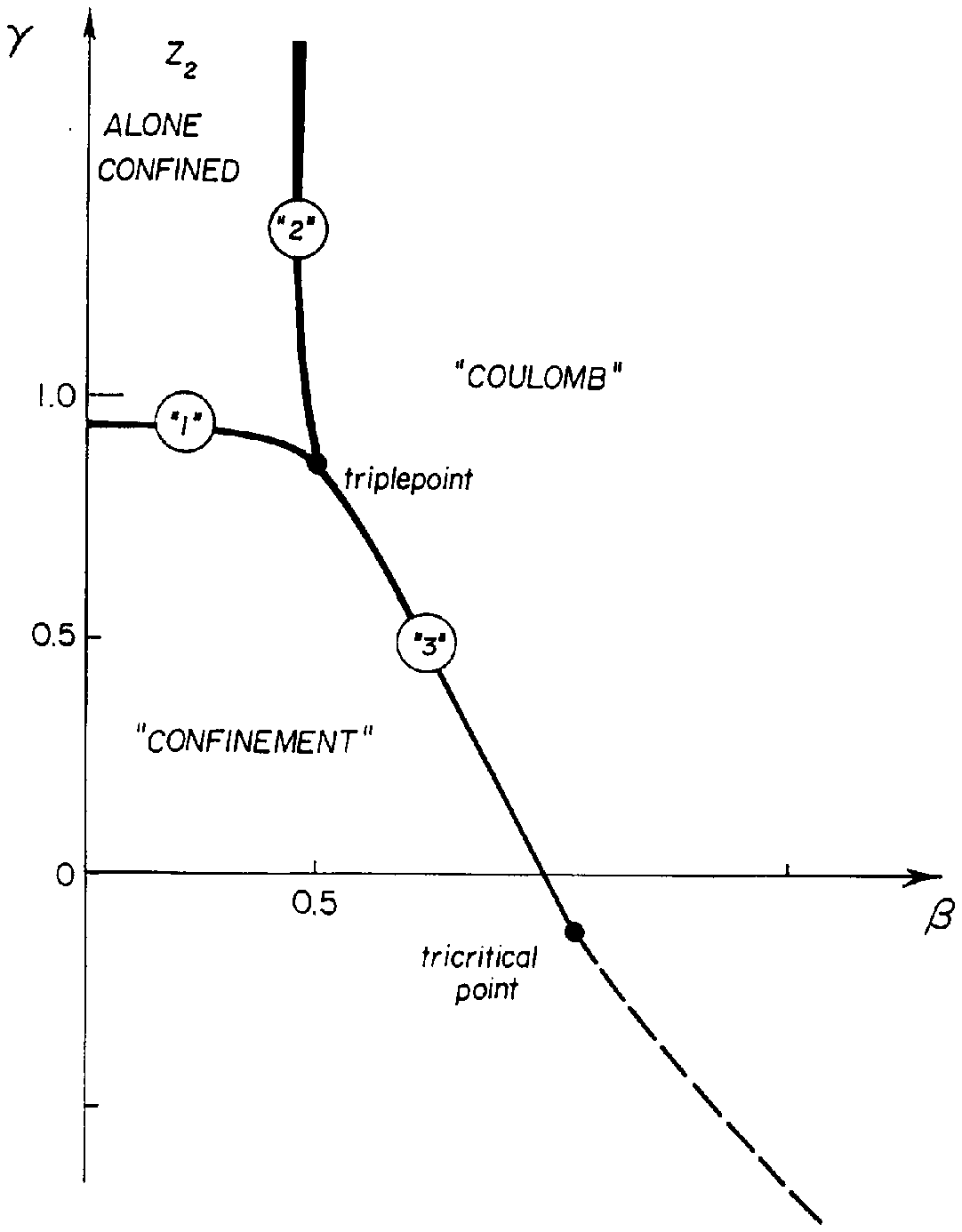}}
\caption{The phase diagram for $U(1)$ when the two-parameter
lattice action is used. This type of action makes it possible to provoke
the confinement $Z_2$ (or $Z_3$) alone. The diagram shows the
existence of a triple (critical) point.
From this triple point emanate three phase borders:
the phase border "1" separates the totally confining phase from the phase
where only the discrete subgroup $Z_{2}$ is confined; the phase border
"2" separates the latter phase from the totally Coulomb--like phase;
and the phase border "3" separates the totally confining and totally
Coulomb--like phases.}
\end{figure}
\newpage

\begin{figure}
\centerline{\epsfxsize=\textwidth \epsfbox{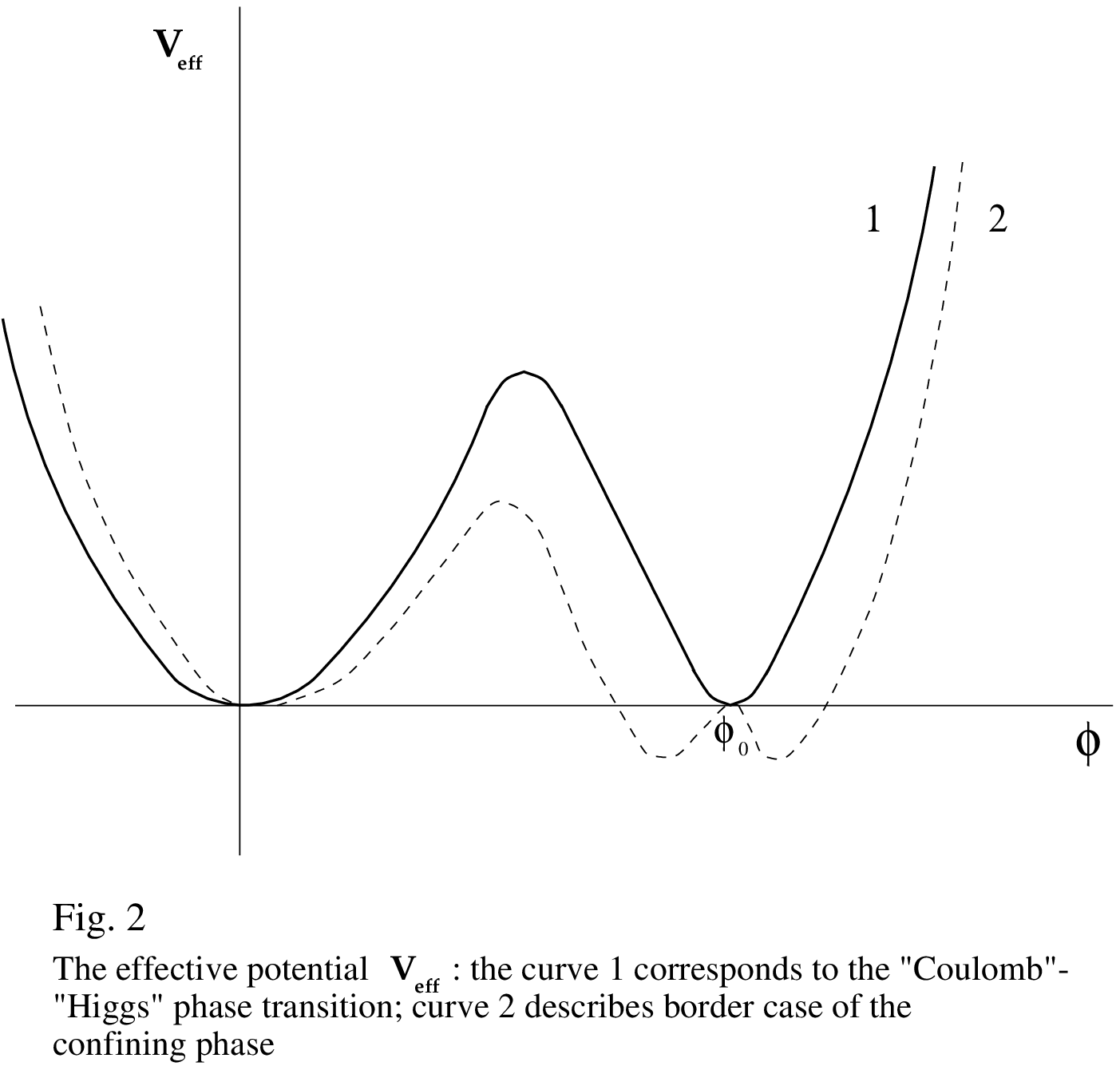}}
\end{figure}

\setcounter{figure}{2}

\newpage

\clearpage

\begin{figure}[t]
\centerline{\epsfxsize=\textwidth \epsfbox{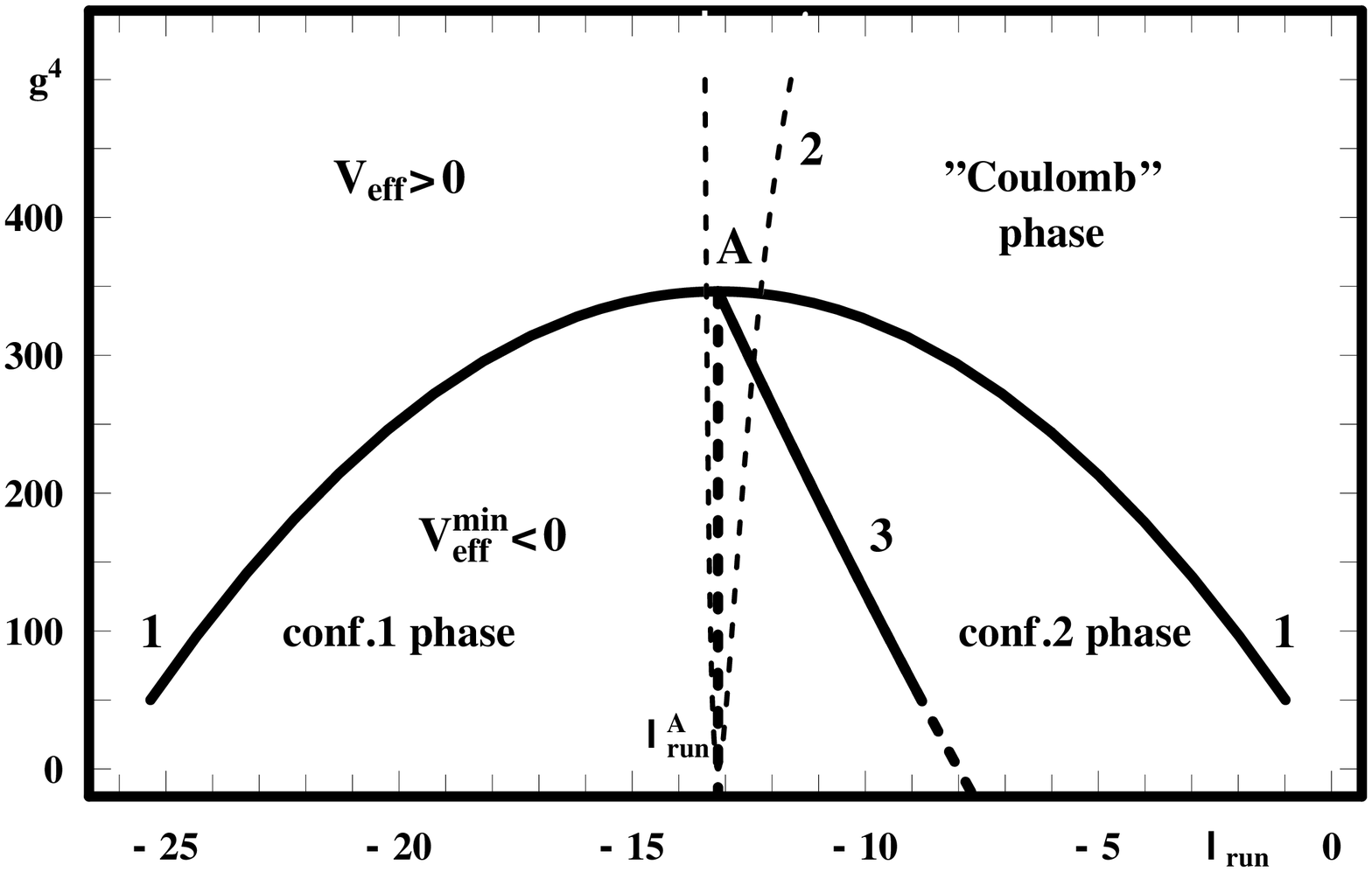}}

\caption{The phase diagram ($\lambda_{run};\; {\rm g}^4\equiv g^4_{run}$)
corresponding to the Higgs monopole model shows the existence of a triple
point A $\bigl(\lambda_{(A)} \approx -13.4;\;{\rm g}^2_{(A)}\approx
18.6\bigr)$. This triple point is a boundary point of three phase
transitions: the "Coulomb-like" phase and two confinement phases
("conf.1" and "conf.2") meet together at the triple point A.
The dashed curve "2" shows the requirement:
$V_{eff}(\phi_0^2) = V''_{eff}(\phi_0^2) = 0$. Monopole condensation
leads to the confinement of the electric charges: ANO electric
vortices (with electric charges at their ends, or closed)
are created in the confinement phases "conf.1" and
"conf.2".}

\end{figure}


\newpage
\begin{thebibliography}{99}
\bibitem{1}
G.Bhanot, Nucl.Phys.{\bf B205}, 168 (1982); Phys.Rev. {\bf D24}, 461 (1981);\\
Nucl.Phys. {\bf B378} 633 (1992).
\bibitem{2}
J.Jersak, T.Neuhaus and P.M.Zerwas, Phys.Lett. {\bf B133} 103 (1983).
\bibitem{3}
J.Jersak, T.Neuhaus and P.M.Zerwas, "Charge renormalization in compact \\
lattice QED", PITHA 84/08, Aachen, Tech. Hochsch., May 1984;\\
published in Nucl.Phys. {\bf B251}, 299 (1985).
\bibitem{4}
H.G.Everetz, T.Jersak, T.Neuhaus, P.M.Zervas, Nucl.Phys. {\bf B251}, 279\\
(1985).
\bibitem{5}
J.Jersak, T.Neuhaus, H.Pfeiffer, "Scaling of Magnetic Monopoles in the \\
Pure Compact QED", The 17th International Symposium on Lattice Field\\
Theory (LATTICE'99); Nucl.Phys.Proc.Suppl. {\bf 83-84}, 491 (2000).
\bibitem{6}
J.Jersak, T.Neuhaus, H.Pfeiffer, Phys.Rev. {\bf D60}, 054502 (1999).
\bibitem{7}
C.P.Bachas and R.F.Dashen, Nucl.Phys. {\bf B210}, 583 (1982).
\bibitem{8}
H.B.Nielsen and D.L.Bennett, "Fitting the Fine Structure Constants by
Critical Couplings and Integers",in: H.J.Kaiser, editor, {\it Proceedings
of the XXV International Symposium Ahrenshoop on the Theory of Elementary
Particles}, page 366. Institut fur Hochenergiephysik, Platanenallee 6,\\
D-O-1615 Zeuthen, Germany, Gosen, Sept.23-26 1991.
\bibitem{9}
D.L.Bennett and H.B.Nielsen, "Standard Model Couplings from Mean Field
Criticality at the Planck Scale and a Maximum Entropy Principle", in:
D.Axen, D.Bryman, and M.Comyn, editors,{\it Proceedings of the \\
Vancouver Meeting on Particles and Fields '91}, 18-22 August,\\
World Scientific Publishing Co., Singapore, 1992, page 857.
\bibitem{10}
D.L.Bennett and H.B.Nielsen, Int. J. Mod. Phys. {\bf A9}, 5155 (1994).
\bibitem{11}
L.V.Laperashvili, Phys.of Atom.Nucl. {\bf 57}, 471 (1994);
\bibitem{12}
L.V.Laperashvili, Phys.of Atom.Nucl. {\bf 59}, 162 (1996).
\bibitem{13}
L.V.Laperashvili, H.B.Nielsen, Mod.Phys.Let. {\bf A12}, 73 (1997).
\bibitem{14}
C.D.Froggatt, L.V.Laperashvili, H.B.Nielsen, "SUSY or NOT SUSY: Anti-GUT's,
Critical Coupling Universality and Higgs--Top Masses", "SUSY98",
Oxford, 10-17 July 1998; hepwww.rl.ac.uk/susy98/.
\bibitem{15}
L.V.Laperashvili, H.B.Nielsen,
"Multiple Point Principle
and Phase Transition in Gauge Theories", in:{\it Proceedings of the
International Workshop on "What Comes Beyond the Standard Model"},
Bled, Slovenia, 29 June - 9 July 1998; Ljubljana 1999, p.15.
\bibitem{16}
L.V.Laperashvili, H.B.Nielsen, "Phase Transition Coupling Constants
in the Higgs Monopole Model", hep-th/9711388.
\bibitem{17}
D.Zwanziger, Phys.Rev. {\bf D3}, 343 (1971).
\bibitem{18}
R.A.Brandt, F.Neri, D.Zwanziger, Phys.Rev.{\bf D19}, 1153 (1979).
\bibitem{19}
F.V.Gubarev, M.I.Polikarpov, V.I.Zakharov, Phys.Lett.{\bf B438}, 147 (1998).
\bibitem{20}
L.V.Laperashvili, H.B.Nielsen, Mod.Phys.Lett. {\bf A14}, 2797 (1999).
\bibitem{21}
S.Coleman, E.Weinberg, Phys.ReV. {\bf D7}, 1888 (1973).
\bibitem{22}
M.Sher, Phys.Rept. {\bf 179}, 274 (1989).
\bibitem{23}
D.Gross, in: Methods of Field Theory, Proc. 1975 Les Houches Summer School,
eds R.Balian and J.Zinn-Justin, North Holland, Amsterdam, 1975.
\bibitem{24}
H.B.Nielsen, "Dual Strings. Fundamental of Quark Models", in: \\
{\it Proceedings of the XVII Scottish University Summer Scool in Physics},\\
St.Andrews, 1976, p.528.
\bibitem{25}
D.L.Bennett, H.B.Nielsen, I.Pi\^cek, Phys.Lett. {\bf B208}, 275 (1988).
\bibitem{26}
H.B.Nielsen, N.Brene, Phys.Lett. {\bf B233}, 399 (1989); Nucl.Phys.
{\bf B224},\\
396 (1983).
\bibitem{27}
C.D.Froggatt, H.B.Nielsen, {\it Origin of Symmetries}, Singapore: World \\
Scientific, 1991.
\bibitem{28}
C.D.Froggatt, M.Gibson, H.B.Nielsen, D.J.Smith, "The Fermion Mass Problem
and the Anti--Grand Unification Model", in: {\it Proceedings of the
29th International Conference on High Energy Physics}, Vancouver,
Canada, 23--29 July, 1998; Int.J.Mod.Phys. {\bf A13}, 5037 (1998).
\bibitem{29}
N.N.Bogoljubov, D.V.Shirkov, Doklady AN SSSR (Reports of AS USSR),\\
{\bf 103}(1955)203; ibid {\bf 103}, 391 (1955); JETP, {\bf 30}, 77 (1956).
\bibitem{30}
L.D.Landau, A.A.Abrikosov, I.M.Khalatnikov, Doklady AN SSSR \\
(Reports of AS USSR), {\bf 95}, 773 (1954); ibid {\bf 95}, 1177 (1954).
\bibitem{31}
P.A.M.Dirac, Proc.Roy.Soc. {\bf A33}, 60 (1931).
\bibitem{32}
J.Shwinger, Phys.Rev. {\bf 44}, 1087 (1996);ibid {\bf 151}, 1048, 1055
(1966);\\
ibid {\bf 173}, 1536 (1968); Science {\bf 165}, 757 (1969); ibid
{\bf 166}, 690 (1969).
\bibitem{33}
M.Blagojevich, P.Senjanovich, Phys.Rept. {\bf 157}, 234 (1988).
\bibitem{34}
H.Alhendi, Phys.Rev. {\bf D37}, 3749 (1988).
\bibitem{35}
H.Arason, D.J.Castano, B.Kesthelyi, S.Mikaelian,
E.J.Piard, P.Ramond, B.D.Wright, Phys.Rev. {\bf D46}, 3945 (1992).
\bibitem{36}
H.B.Nielsen, P.Olesen, Nucl.Phys., {\bf B61}, 45 (1973).
\bibitem{37}
A.A.Abrikosov, Soviet JETP, {\bf 32}, 1442 (1957).
\bibitem{38}
E.T.Akhmedov, M.N.Chernodub, M.I.Polikarpov, M.A.Zubkov,
Phys.Rev. {\bf D53}, 2087 (1996).
\bibitem{39}
M.L\"uscher, K.Symanzik, P.Weisz, Nucl.Phys. {\bf B173}, 365 (1980).
\bibitem{40}
H.P.Nilles, Phys.Reports {\bf 110}, 1 (1984).
\bibitem{41}
P.Langacker, N.Polonsky, Phys.Rep. {\bf D47}, 4028 (1993).
\bibitem{42}
K.A.Olive, "Introduction to Supersymmetry: Astrophysical and\\
Phenomenological Constraints", hep-ph/9911307.
\bibitem{43}
Gi-Chol Cho, Kaoru Hagiwara, "Supersymmetry versus precision \\
experiments revisted", hep-ph/9912260.

\end{thebibliography}
\end{document}